\documentstyle[preprint,prb,aps]{revtex}

\begin{document}

\begin{titlepage}

\title
{Spin-phase interference, coherent superposition, and quantum tunneling at
excited levels in nano-antiferromagnets}

\author{Rong L\"{u}\footnote {Author to whom 
the correspondence should be addressed.\\
Electronic address: rlu@castu.tsinghua.edu.cn}, Jia-Lin Zhu, Yi Zhou, and Bing-Lin Gu} 
\address{Center for Advanced Study, 
Tsinghua University, Beijing 100084, People's Republic of China}
\date{\today}

\maketitle
\begin{abstract}
The spin-phase interference effects
are studied analytically in resonant quantum tunneling
of the N\'{e}el  vector between degenerate excited levels
in nanometer-scale single-domain antiferromagnets in the absence of an external
magnetic field.
We consider a model for mesoscopic antiferromagnets with uncompensated
excess spins for the more general structure of magnetic anisotropy,
such as biaxial, trigonal, tetragonal and hexagonal crystal symmetry.
This study provides a nontrivial generalization of the Kramers degeneracy
for double-well system to coherently spin tunneling at ground states
as well as low-lying excited states in AFM
system with $m$-fold rotational symmetry around the $\widehat{z}$ axis.
The energy level spectrum and the thermodynamic properties of magnetic tunneling states
are found to depend significantly on the parity of the excess spins
at sufficiently low temperatures. Possible relevance to experiments is also
discussed.

\noindent
{\bf PACS number(s)}:  75.45.+j, 75.50.Ee, 75.10.Jm, 03.65.Bz
\end{abstract}

\end{titlepage}

\section{Introduction}

Can quantum mechanics be used to describe the behavior of macroscopic
objects? This question has fascinated physicists for more than 70 years
until the 1980s when it was proposed by Leggett {\it et al}.\cite{Leggett}
that macroscopic objects could behave quantum mechanically, provided that
the dissipative interactions with the environment were small enough. In
recent years, macroscopic quantum phenomena have been observed in various
systems---for example, quantum tunneling of the phase in a Josephson
junction, permanent current in small conductor rings, Bose-Einstein
condensation in atomic vapours, and C$_{60}$ molecules. Recently, tunneling
of the magnetization has been intensively studied theoretically and
experimentally in nanoparticles and molecular clusters.\cite{review} Two
kinds of phenomena can be envisaged. First, macroscopic quantum tunneling
(MQT), where the magnetization tunnels through an barrier from one
metastable state to a stable one. Second, macroscopic quantum coherence
(MQC), where the magnetization oscillates back and forth between the
degenerate states. This oscillation should show up in the
frequency-dependent magnetic noise and susceptibility spectra.\cite
{review,Awschalom} Besides its importance from a fundamental point of view,
tunneling of the magnetization changes the properties of small magnets, with
potential implications for the data-storage technology. It is also very
important to the reliability of nanometer-scale magnetic units in memory
devices and the designing of quantum computers.\cite{Tejade}

One notable subject in spin quantum coherence is that the topological
Wess-Zumino-Berry phase\cite{Fradkin} can lead to remarkable spin-parity
effects. It was found that the tunnel splitting is suppressed to zero for
half-integer total spins in biaxial ferromagnetic (FM) particles due to the
destructive phase interference between topologically different tunneling
paths.\cite{Loss(92) and Delft (92)} However, the phase interference is
constructive for integer spins, and hence the splitting is nonzero.\cite
{Loss(92) and Delft (92)} While spin-parity effects are sometimes be related
to Kramers degeneracy, they typically go beyond the Kramers theorem in a
rather unexpected way.\cite{Garg (93) and Loss (95)(96)} The auxiliary
particle method was proposed to study the model for a single large spin
subject to the external and anisotropy fields, and to discuss the
spin-parity effects.\cite{Barnes} Similar effect was found in
antiferromagnetic (AFM) particles, where only the integer excess spins can
tunnel but not the half-integer ones.\cite{Bogachek and Krive,AFM} However,
these spin-parity effects are intrinsically absent in the phenomena of MQT
and MQC in the Josephson-junction-based superconducting systems, which makes
the tunneling phenomena in nanoscale magnets more important for
understanding the foundations of quantum mechanics. Theoretical results
showed that MQC in AFM particles should show up at higher temperatures and
higher frequencies than in FM particles of similar size,\cite{Bogachek and
Krive,AFM,Levine,Loss (97)(98)} which makes AFM particles more interesting
for experimental purposes. Recently, topological phase interference effects
were investigated extensively in FM and AFM particles in a magnetic field,%
\cite{Garg (93) and Loss (95)(96),Barnes,Bogachek and
Krive,Chudnovsky93,Garg95,Wang,Lu1,Liang1} and in the systems with different
symmetries.\cite{Wang2,Lu2}

One recent experiment\cite{Wernsdorfer} was performed to measure the tunnel
splittings in molecules Fe$_8$, and a clear oscillation of the splitting as
a function of the field along the hard axis was observed, which is a direct
evidence of the role of the topological spin phase (Berry phase) in the spin
dynamics. Recent theoretical and experimental studies include the thermally
activated resonant tunneling based on the exact diagonalization,\cite
{Garanin} the auxiliary particle method,\cite{Barnes} the discrete WKB\
method and a nonperturbation calculation,\cite{Garg} the non-adiabatic
Landau-Zener model,\cite{Chiorescu} and the calculation based on exact
spin-coordinate correspondence.\cite{Liang2}

The importance of the topological interference term of the Berry phase for
the problem of spin tunneling and the associated spin-parity effects have
been elucidated in Refs. 6, 7 and 10. However, the theoretical studies on
AFM systems\cite{Bogachek and Krive,AFM,Levine,Loss (97)(98)} have been
focused on phase interference between two opposite winding ground-state
tunneling paths in biaxial particles. The spin-phase interference between
excited-level tunneling paths is not clearly shown for AFM particles.
Moreover, the previous works on AFM spin tunneling\cite{Bogachek and
Krive,AFM,Levine,Loss (97)(98)} have been confined to the system with
biaxial symmetry, which has two energetically degenerate easy directions in
the basal plane. The purpose of this paper is to study the quantum tunneling
and spin-phase interference {\it at excited states} for AFM particles in the
absence of an external magnetic field. By calculating the nonvacuum
instantons, we obtain the analytical results for tunnel splittings at
excited levels. To compare theory with experiment, we consider the AFM
particles with {\it the general structure of magnetocrystalline anisotropy},
such as biaxial, trigonal, tetragonal, and hexagonal symmetry around $%
\widehat{z}$, which have two, three, four, and six energetically degenerate
easy directions in the basal plane. For AFM particles with biaxial symmetry,
the spin-phase interference effect can be studied by summing up the
contributions of topologically different tunneling paths of clockwise and
counterclockwise instantons.\cite{Bogachek and Krive,AFM,Levine,Loss
(97)(98)} However, for the system with complex trigonal, tetragonal or
hexagonal symmetry, this procedure is hard to evaluate. In this paper, the
spin tunneling problem is mapped onto a particle moving problem in
one-dimensional periodic potential $U\left( \phi \right) $ by integrating
out the momentum in the path integral, and the tunneling level spectrum of
excited states is obtained by using the Bloch theorem. Our results show that
the {\it excited-level} tunnel splittings depend significantly on the parity
of the excess spins of AFM particles. The low-energy limits of the nonvacuum
instanton and the tunnel splittings of excited levels agree well with the
results of ground-state tunneling in biaxial AFM particles.\cite{AFM} The
structure of tunneling level spectrum for the trigonal, tetragonal and
hexagonal crystal symmetry is found to be much more complex than that for
the biaxial crystal symmetry.\cite{AFM,Levine,Loss (97)(98)} The tunnel
splitting for biaxial AFM particles is quenched to zero for half-integer
excess spins due to the destructive interference of Berry phase.\cite
{Loss(92) and Delft (92),Bogachek and Krive,AFM,Levine,Loss (97)(98)} While
the tunnel splitting can be nonzero even if the excess spin is a
half-integer for the trigonal, tetragonal, or hexagonal symmetry at zero
magnetic field.

This paper is structured in the following way. In Sec. II, we review briefly
some basic ideas of quantum tunneling in AFM\ particles, and discuss the
fundamentals concerning the computation of excited-level splittings in the
double-well potential. In Secs. III we study the resonant quantum tunneling
of N\'{e}el vector between degenerate excited states in AFM particles with
the biaxial symmetry in detail, and present the results for trigonal,
tetragonal and hexagonal symmetry in Sec. IV. The conclusions are presented
in Sec. V.

\section{Physical model}

The system of interest is a single-domain AFM particle of about 5$\sim $10
nm in radius at a temperature well below its anisotropy gap. According to
the two-sublattice model,\cite{AFM} there is a strong exchange energy ${\bf m%
}_1\cdot {\bf m}_2/\chi _{\bot }$ between two sublattices, where ${\bf m}_1$
and ${\bf m}_2$ are the magnetization vectors of the two sublattices with
large, fixed and unequal magnitudes, and $\chi _{\bot }$ is the transverse
susceptibility. Under the assumption that the exchange energy between two
sublattices is much larger than the magnetocrystalline anisotropy energy,
the Euclidean action for the AFM particle (neglecting dissipation with the
environment) is\cite{Bogachek and Krive,AFM,Levine,Loss (97)(98)} 
\begin{eqnarray}
{\cal S}_E[\theta ({\bf x},\tau ),\phi ({\bf x},\tau )] &=&\frac 1\hbar \int
d\tau \int d^3x\left\{ i\frac{m_1+m_2}\gamma \left( \frac{d\phi }{d\tau }%
\right) +\frac m\gamma \left( \frac{d\phi }{d\tau }\right) \cos \theta +%
\frac{\chi _{\bot }}{2\gamma ^2}\left[ \left( \frac{d\theta }{d\tau }\right)
^2\right. \right.  \nonumber \\
&&\left. \left. +\left( \frac{d\phi }{d\tau }\right) ^2\sin ^2\theta \right]
+\frac 12\alpha \left[ (\nabla \theta )^2+(\nabla \phi )^2\sin ^2\theta
\right] +E(\theta ,\phi )\right\} ,  \eqnum{1}
\end{eqnarray}
where $\gamma $ is the gyromagnetic ratio, $\alpha $ is the exchange
constant (which is also referred to as the stiffness constant, or the Bloch
wall coefficient\cite{Kittle}), and $\tau =it$ is the imaginary-time
variable. The $E(\theta ,\phi )$ term includes the magnetocrystalline
anisotropy and the Zeeman energies. The polar coordinate $\theta $ and the
azimuthal coordinate $\phi $, which are the angular components of ${\bf m}_1$
in the spherical coordinate system, can determine the direction of the
N\'{e}el vector.

As pointed out in Ref. 10, for a nanometer-scale single-domain AFM particle,
the N\'{e}el vector may depend on the imaginary time but not on coordinates
because the spatial derivatives in Eq. (1) are suppressed by the strong
exchange interaction between two sublattices. So all the calculations
performed in the present work are for the homogeneous N\'{e}el vector.
Therefore, Eq. (1) reduces to 
\begin{eqnarray}
{\cal S}_E(\theta ,\phi ) &=&\frac V\hbar \int d\tau \left\{ i\frac{m_1+m_2}%
\gamma \left( \frac{d\phi }{d\tau }\right) +\frac m\gamma \left( \frac{d\phi 
}{d\tau }\right) \cos \theta \right.  \nonumber \\
&&\left. +\frac{\chi _{\bot }}{2\gamma ^2}\left[ \left( \frac{d\theta }{%
d\tau }\right) ^2+\left( \frac{d\phi }{d\tau }\right) ^2\sin ^2\theta
\right] +E\left( \theta ,\phi \right) \right\} ,  \eqnum{2}
\end{eqnarray}
where $V$ is the volume of the single-domain AFM nanoparticle. $%
m=m_1-m_2=\hbar \gamma s/V$, where $s$ is the excess spin due to the
noncompensation of two sublattices. Note that the first term in the
Euclidean action is a total imaginary-time derivative. Its integral depends
only on the initial and final states and hence has no effect on the
classical equations of motion, but yields a boundary contribution to the
Euclidean action. However, it was shown that this term, known as the
topological phase term, is of central importance for the quantum
interference effect and makes the tunneling behaviors of integer and
half-integer excess spins strikingly different.\cite{Bogachek and
Krive,AFM,Levine,Loss (97)(98)}

The Euclidean transition amplitude from an initial state $|\theta _i,\phi
_i\rangle $ to a final state $|\theta _f,\phi _f\rangle $ can be expressed
as the following imaginary-time path integral in the spin-coherent-state
representation, 
\begin{equation}
\langle \theta _f,\phi _f|e^{-{\cal H}T}|\theta _i,\phi _i\rangle =\int 
{\cal D}\{\theta \}{\cal D}\{\phi \}\exp [-{\cal S}_E(\theta ,\phi )], 
\eqnum{3}
\end{equation}
where the Euclidean action ${\cal S}_E(\theta ,\phi )$ has been defined in
Eq. (2). In the semiclassical limit, the dominant contribution to the
transition amplitude comes from finite action solutions of the classical
equations of motion (instantons). According to the standard instanton
technique, the tunneling rate $\Gamma $ for MQT or the tunnel splitting $%
\Delta $ for MQC is given by $\Gamma ($or $\Delta )=Ae^{-{\cal S}_{cl}}$,%
\cite{Coleman} where ${\cal S}_{cl}$ is the WKB exponent or the classical
action which minimizes the Euclidean action of Eq. (2). The preexponential
factor $A$ originates from the quantum fluctuations about the classical
path, which can be evaluated by expanding the Euclidean action to the second
order in small fluctuations.\cite{Coleman} It is noted that the above result
is based on tunneling at the ground state, and the temperature dependence of
the tunneling frequency (i.e., tunneling at excited states) is not taken
into account. The instanton technique is suitable only for the evaluation of
the tunneling rate at the vacuum level, since the usual (vacuum) instantons
satisfy the vacuum boundary conditions. Different types of pseudoparticle
configurations were developed which satisfy periodic boundary condition
(i.e., periodic instantons or nonvacuum instantons).\cite{Liang3}

For a particle moving in a double-well-like potential $U\left( x\right) $,
the WKB method gives the tunnel splitting of $n$th excited states as\cite
{Landau} 
\begin{equation}
\Delta E_n=\frac{\omega \left( E_n\right) }\pi \exp \left[ -{\cal S}\left(
E_n\right) \right] ,  \eqnum{4}
\end{equation}
with the imaginary-time action is 
\begin{equation}
{\cal S}\left( E_n\right) =\sqrt{2m}\int_{x_1\left( E_n\right) }^{x_2\left(
E_n\right) }dx\sqrt{U\left( x\right) -E_n},  \eqnum{5}
\end{equation}
where $x_{1,2}\left( E_n\right) $ are the turning points for the particle
oscillating in the inverted potential $-U\left( x\right) $. $\omega \left(
E_n\right) =2\pi /t\left( E_n\right) $ is the energy-dependent frequency,
and $t\left( E_n\right) $ is the period of the real-time oscillation in the
potential well, 
\begin{equation}
t\left( E_n\right) =\sqrt{2m}\int_{x_3\left( E_n\right) }^{x_4\left(
E_n\right) }\frac{dx}{\sqrt{E_n-U\left( x\right) }},  \eqnum{6}
\end{equation}
where $x_{3,4}\left( E_n\right) $ are the classical turning points for the
particle oscillating inside $U\left( x\right) $. The functional-integral and
the WKB\ method showed that for the potentials parabolic near the bottom the
result Eq. (4) should be multiplied by $\sqrt{\frac \pi e}\frac{\left(
2n+1\right) ^{n+1/2}}{2^ne^nn!}$.\cite{Garanin2,Weiss} This factor is very
close to 1 for all $n$: 1.075 for $n=0$, 1.028 for $n=1$, 1.017 for $n=2$,
etc. Stirling's formula for $n!$ shows that this factor trends to 1 as $%
n\rightarrow \infty $. Therefore, this correction factor, however, does not
change much in front of the exponentially small action term in Eq. (4).
Recently, the crossover from quantum to classical behavior and the
associated phase transition have been investigated extensively in nanospin
systems.\cite{Garanin2,Chudnovsky(PRL97),Liang4,Gorokhov,Park}

\section{MQC for biaxial symmetry}

In this section, we consider an AFM system with biaxial symmetry, i.e.,
which has two degenerate easy directions in the biaxial plane. Now the
magnetocrystalline anisotropy energy is 
\begin{equation}
E\left( \theta ,\phi \right) =K_1\cos ^2\theta +K_2\sin ^2\theta \sin ^2\phi
+E_0,  \eqnum{7}
\end{equation}
where $K_1$ and $K_2$ are the transverse and longitudinal anisotropic
constants satisfying $K_1\gg K_2>0$, and $E_0$ is a constant which makes $%
E\left( \theta ,\phi \right) $ zero at the initial state. As $K_1\gg K_2>0$,
the N\'{e}el vector is forced to lie in the $\theta =\pi /2$ plane, so the
fluctuations of $\theta $ about $\pi /2$ are small. Introducing $\theta =\pi
/2+\alpha $ $\left( \left| \alpha \right| \ll 1\right) $, Eq. (7) reduces to 
\begin{equation}
E\left( \alpha ,\phi \right) \approx K_1\alpha ^2+K_2\sin ^2\phi .  \eqnum{8}
\end{equation}
The ground state corresponds to the N\'{e}el vector pointing in one of the
two degenerate easy directions: $\theta =\pi /2$, and $\phi =0$, $\pi $,
other energy minima repeat the two states with period $2\pi $. Performing
the Gaussian integration over $\alpha $, we can map the spin system onto a
particle moving problem in one-dimensional potential well. Now the
transition amplitude becomes 
\begin{eqnarray}
U_{fi} &=&\exp \left[ -iS_{tot}\left( \phi _f-\phi _i\right) \right] \int
d\phi \exp \left( -{\cal S}_E\left[ \phi \right] \right) ,  \nonumber \\
&=&\exp \left[ -iS_{tot}\left( \phi _f-\phi _i\right) \right] \int d\phi
\exp \left\{ -\int d\tau \left[ \frac 12M\left( \frac{d\phi }{d\tau }\right)
^2+U\left( \phi \right) \right] \right\} ,  \eqnum{9}
\end{eqnarray}
with $S_{tot}=2S-s$ being the total spins of two sublattices, $M=\frac V\hbar
\left( \frac{\chi _{\bot }}{\gamma ^2}+\frac{m^2}{2K_1\gamma ^2}\right) =%
\frac{\hbar S^2}{JV}\left[ 1+\frac 12\left( \frac J{K_1}\right) \left( \frac %
sS\right) ^2\right] $ being the effective mass, and $U\left( \phi \right)
=\left( K_2V/\hbar \right) \sin ^2\phi $ being the effective potential. $J$
is the exchange density between two sublattices, which is related to the
transverse susceptibility $\chi _{\bot }$ by taking the simple estimate as $%
\chi _{\bot }\approx \hbar ^2\gamma ^2S^2/JV^2$,\cite{AFM} $S=m_1V/\hbar
\gamma $ is the total spin in ${\bf m}_1$ sublattice, and $s=mV/\hbar \gamma 
$ is the excess spin due to the noncompensation of two sublattices. It is
noted that the total derivative in Eq. (2), when integrated, gives an
additional phase factor to the transition amplitude Eq. (9) which depends on
the initial and final values of $\phi $. For the biaxial symmetry, this
phase factor in Eq. (9) is $\exp \left( -i\pi S_{tot}\right) $. The
potential $U\left( \phi \right) $ is periodic with period $\pi $, and there
are two minima in the entire region $2\pi $. We may look at $U\left( \phi
\right) $ as a superlattice with lattice constant $\pi $ and total length $%
2\pi $, and we can derive the energy spectrum by applying the Bloch theorem
and the tight-binding approximation. The translational symmetry is ensured
by the possibility of successive $2\pi $ extensions.

The periodic instanton configuration $\phi _p$ which minimizes the Euclidean
action in Eq. (9) satisfies the equation of motion 
\begin{equation}
\frac 12M\left( \frac{d\phi _p}{d\tau }\right) ^2-U\left( \phi _p\right) =-E,
\eqnum{10}
\end{equation}
where $E>0$ is a constant of integration, which can be viewed as the
classical energy of the pseudoparticle configuration. Then we obtain the
kink-solution as 
\begin{equation}
\sin ^2\phi _p=1-k^2\text{sn}^2\left( \omega _1\tau ,k\right) ,  \eqnum{11}
\end{equation}
where sn$\left( \omega _1\tau ,k\right) $ is the Jacobian elliptic sine
function of modulus $k$, 
\begin{equation}
k^2=\frac{n_1^2-1}{n_1^2},  \eqnum{12}
\end{equation}
with $\omega _1=\sqrt{2}\frac V{\hbar S}\sqrt{\frac{K_2J}{1+\frac 12\left( 
\frac J{K_1}\right) \left( \frac sS\right) ^2}}$, and $n_1=\sqrt{K_2V/\hbar E%
}>1$. In the low energy limit, i.e., $E\rightarrow 0$, $k\rightarrow 1$, sn$%
\left( u,1\right) \rightarrow \tanh u$, we have 
\begin{equation}
\sin \phi _p=\frac 1{\cosh \left( \omega _1\tau \right) },  \eqnum{13}
\end{equation}
which is exactly the vacuum instanton solution derived in Ref. 10.

The Euclidean action of the periodic instanton configuration Eq. (11) over
the domain $\left( -\beta ,\beta \right) $ is found to be 
\begin{equation}
{\cal S}_p=\int_{-\beta }^\beta d\tau \left[ \frac 12M\left( \frac{d\phi _p}{%
d\tau }\right) ^2+V\left( \phi _p\right) \right] =W+2E\beta ,  \eqnum{14}
\end{equation}
with 
\begin{equation}
W=2^{3/2}S\sqrt{\frac{K_2}J\left[ 1+\frac 12\left( \frac J{K_1}\right)
\left( \frac sS\right) ^2\right] }\left[ E\left( k\right) -\left(
1-k^2\right) K\left( k\right) \right] ,  \eqnum{15}
\end{equation}
where $K\left( k\right) $ and $E\left( k\right) $ are the complete elliptic
integral of the first and second kind, respectively. Now we discuss the low
energy limit where $E$ is much less than the barrier height. In this case, $%
k^{\prime 2}=1-k^2=\hbar E/K_2V\ll 1$, so we can perform the expansions of $%
K\left( k\right) $ and $E\left( k\right) $ in Eq. (15) to include terms like 
$k^{\prime 2}$ and $k^{\prime 2}\ln \left( 4/k^{\prime }\right) $, 
\begin{eqnarray}
E\left( k\right) &=&1+\frac 12\left[ \ln \left( \frac 4{k^{\prime }}\right) -%
\frac 12\right] k^{\prime 2}+\cdots ,  \nonumber \\
K\left( k\right) &=&\ln \left( \frac 4{k^{\prime }}\right) +\frac 14\left[
\ln \left( \frac 4{k^{\prime }}\right) -1\right] k^{\prime 2}+\cdots . 
\eqnum{16}
\end{eqnarray}
With the help of small oscillator approximation for energy near the bottom
of the potential well, $E={\cal E}_n^{bia}=\left( n+1/2\right) \omega _1$,
Eq. (15) is expanded as 
\begin{eqnarray}
W &=&2^{3/2}S\sqrt{\frac{K_2}J\left[ 1+\frac 12\left( \frac J{K_1}\right)
\left( \frac sS\right) ^2\right] }-\left( n+\frac 12\right)  \nonumber \\
&&+\left( n+\frac 12\right) \ln \left[ \frac{\left( n+\frac 12\right) }{%
2^{7/2}}\frac 1{S\sqrt{\frac{K_2}J\left[ 1+\frac 12\left( \frac J{K_1}%
\right) \left( \frac sS\right) ^2\right] }}\right] .  \eqnum{17}
\end{eqnarray}
Then the general formula Eq. (4) gives the low-lying energy shift of $n$th
excited states for AFM particles with biaxial crystal symmetry at zero
magnetic field as 
\begin{eqnarray}
\hbar \Delta {\cal E}_n^{bia} &=&\frac 1{\sqrt{2\pi }n!}\frac{\sqrt{K_2J}V}{S%
\sqrt{1+\frac 12\left( \frac J{K_1}\right) \left( \frac sS\right) ^2}}\left(
2^{7/2}S\sqrt{\frac{K_2}J\left[ 1+\frac 12\left( \frac J{K_1}\right) \left( 
\frac sS\right) ^2\right] }\right) ^{n+1/2}  \nonumber \\
&&\times \exp \left( -2^{3/2}S\sqrt{\frac{K_2}J\left[ 1+\frac 12\left( \frac %
J{K_1}\right) \left( \frac sS\right) ^2\right] }\right) .  \eqnum{18}
\end{eqnarray}

When $n=0$, the energy shift of the ground state is 
\begin{eqnarray}
\hbar \Delta {\cal E}_0^{bia} &=&\frac{2^{5/4}}{\sqrt{\pi }}\left( \sqrt{K_2J%
}V\right) \left( \frac{K_2}{J\left[ 1+\frac 12\left( \frac J{K_1}\right)
\left( \frac sS\right) ^2\right] }\right) ^{1/4}S^{-1/2}  \nonumber \\
&&\times \exp \left( -2^{3/2}S\sqrt{\frac{K_2}J\left[ 1+\frac 12\left( \frac %
J{K_1}\right) \left( \frac sS\right) ^2\right] }\right) .  \eqnum{19}
\end{eqnarray}
Then Eq. (18) can be written as 
\begin{equation}
\hbar \Delta {\cal E}_n^{bia}=\frac{q^n}{n!}\left( \hbar \Delta {\cal E}%
_0^{bia}\right) ,  \eqnum{20}
\end{equation}
where 
\begin{equation}
q=2^{7/2}S\sqrt{\frac{K_2}J\left[ 1+\frac 12\left( \frac J{K_1}\right)
\left( \frac sS\right) ^2\right] }.  \eqnum{21}
\end{equation}

Now we discuss briefly the dissipation effect and the temperature dependence
of the decay rate. It is noted that Eqs. (20) and (21) are obtained under
the condition that the levels in the two wells are degenerate. In more
general cases, the transition amplitude between two levels separated by the
barrier or the decay rate should be sensitive to this resonance condition
for the two levels. For a spin tunneling problem, it is important to
consider the discrete level structure. It was quantitatively shown that the
phenomenon of MQC depends curcially on the width of the excited levels in
the right well.\cite{Garg95} Including the effects of dissipation, the decay
rate, in particular, is given by\cite{Garg95,Weiss(book)} 
\begin{equation}
\Gamma _n=\frac 12\left( \Delta {\cal E}_n\right) ^2\sum_{n^{\prime }}\frac{%
\Omega _{nn^{\prime }}}{\left( {\cal E}_n-{\cal E}_{n^{\prime }}\right)
^2+\Omega _{nn^{\prime }}^2},  \eqnum{22}
\end{equation}
where $\Delta {\cal E}_n$ is the level splitting, $n^{\prime }$ are the
levels in the other well and $\Omega _{nn^{\prime }}$ is the sum of the
linewidths of the $n$th and $n^{\prime }$th levels caused by the couplying
of the system to the environment. For the exact resonance conditions, the
temperature dependence of the decay rate is 
\begin{equation}
\Gamma \left( T\right) =\sum_n\frac{\left( \Delta {\cal E}_n\right) ^2}{%
2\Omega _n}\exp \left( -{\cal E}_n\beta \right) ,  \eqnum{23}
\end{equation}
where the level broadening $\Omega _n$ contains all the details of the
coupling between the magnet and its environment. If the width caused by the
coupling of the system to the environment is sufficiently large, the levels
overlap, so that the problem is more or less equivalent to the tunneling
into the structureless continuum.\cite{Garg95} In this case, the results
obtained in this paper should be changed by including the dissipation. It is
noted that the purpose of this paper is to study the coherently quantum
tunneling and spin-phase interference at excited levels for AFM particles in
the absence of a magnetic field at sufficiently low temperatures. Strong
dissipation is hardly the case for magnetic systems,\cite{spin-dissipation}
and thereby our results are expected to hold. It has been argued that the
decay rate should oscillate on the applied magnetic field depending on the
relative magnitude between the width and the level spacing.\cite{Garg (93)
and Loss (95)(96),Chudnovsky93,Garg95} However, it is not clear, to our
knowedge, what should be the effect of finite temperature in the problem of
spin tunneling. The full analysis of spin tunneling onto the precession
levels remains an open problem.

Now we consider the transition exponent which is usually addressed by
experiments. Transitions between two states in a bistable system or escaping
from a metastable state can occur either due to the quantum tunneling or via
the classical thermal activation. In the limit of $T\rightarrow 0$, the
transitions are purely quantum-mechanical and the rate goes as $\Gamma \sim
\exp \left( -S_{cl}\right) $, with $S_{cl}$ being the classical action or
the WKB\ exponent which is independent of temperature. As the temperature
increases from zero, thermal effects enter in the quantum tunneling process.
If the temperature is sufficiently high, the decay from a metastable state
is determined by processes of thermal activation, and the transition rate
follows the Arrhenius law, $\Gamma \sim \exp \left( -U/k_BT\right) $, with $%
k_B$ being the Boltzmann constant and $U$ being the height of energy barrier
between the two states. Because of the exponential dependence of the thermal
rate on $T$, the temperature $T_c$ characterizing the crossover from quantum
to thermal regime can be estimated as $k_BT_c=U/S_{cl}$. For the present
case, one can estimate that 
\[
k_BT_c=\frac 1{2\sqrt{2}}\left( \sqrt{K_2J}V\right) S^{-1}\frac 1{\sqrt{1+%
\frac 12\left( \frac J{K_1}\right) \left( \frac sS\right) ^2}}, 
\]
characterizing the crossover from quantum to classical regime. Typical
values of parameters for single-domain AFM nanoparticles are: the sublattice
spin $S=5000$, the excess spin $s=150$, the longitudinal anisotropy constant 
$K_2\sim 10^5$ erg/cm$^3$, the transverse anisotropy constant $K_1\sim 10^6$
erg/cm$^3$, the exchange energy density between two sublattices $J\sim 10^9$
erg/cm$^3$, and the radius of particle is about 5 nm. By taking these
values, we obtain that $T_c\approx $704 mK, which agrees well with the
experimental result for the diluted samples of horse-spleen ferritin.\cite
{review}

It is noted that $\hbar \Delta {\cal E}_n^{bia}$ is only the level shift
induced by tunneling between degenerate excited states through a single
barrier. The effective periodic potential $U\left( \phi \right) =U\left(
\phi +n\pi \right) $ can be regarded as a one-dimensional superlattice with
lattice constant $\pi $. The tunneling through one barrier leads to the
level splitting which extends formally to an energy ``band'' by translation
symmetry ( the rotation symmetry in the present case, $U\left( \phi \right)
=U\left( \phi +n\pi \right) $). The energy ``band'' structure of this
problem is formally the same as that of a one-dimensional tight-binding
model in solid state physics. Then the energy spectrum of low-lying excited
levels can be determined by the Bloch theorem. It is easy to show that if $%
{\cal E}_n^{bia}$ are the degenerate eigenvalues of the system with
infinitely high barrier, the energy level spectrum is given by the following
formula with the help of tight-binding approximation, 
\begin{equation}
E_n^{bia}={\cal E}_n^{bia}-2\Delta {\cal E}_n^{bia}\cos \left[ \left(
S_{tot}+\xi \right) \pi \right] .  \eqnum{24}
\end{equation}
The Bloch wave vector $\xi $ can be assumed to take either of the two values 
$0$ and $1$ in the first Brillouin zone. It is noted that in Eq. (24) we
have included the contribution of {\it topological phase} for AFM\ particles
with biaxial crystal symmetry (i.e., $\pi S_{tot}$). One can easily show
that the low-lying tunneling level spectrum, which corresponds to the
splittings of $n$th {\it excited state} due to the resonant quantum
coherence of the N\'{e}el vector between energetically degenerate states,
depends on the parity of $S_{tot}$ (or the excess spin $s$) significantly.

At the end of this section, we discuss the possible relevance to the
experimental test for spin-parity effects in single-domain AFM
nanoparticles. One can easily show that the specific heat for integer excess
spins is much different from that for half-integer excess spins at
sufficiently low temperatures $T\sim T_0=\hbar \Delta {\cal E}_0^{bia}/k_B$.
When the temperature is higher $\hbar \Delta {\cal E}_0\ll k_BT<\hbar \omega
_1$, the excited energy levels may give contribution to the partition
function of tunneling states. Now the partition function without the
dissipation is given by 
\begin{equation}
{\cal Z}\approx {\cal Z}_0\left[ 1+\frac 12\left( 1-e^{-\beta \hbar \omega
_1}\right) \left( \beta \hbar \Delta \varepsilon _0^{bia}\right) ^2I_0\left(
2qe^{-\beta \hbar \omega _1/2}\right) \right] ,  \eqnum{25}
\end{equation}
for both integer and half-integer excess spins. ${\cal Z}_0=2e^{-\beta \hbar
\omega _1/2}/\left( 1-e^{-\beta \hbar \omega _1}\right) $ is the partition
function in the well calculated for $k_BT\ll \Delta U$ over the low-lying
oscillator like states with ${\cal E}_n^{bia}=\left( n+1/2\right) \omega _1$%
. $I_0\left( x\right) =\sum_{n=0}\left( x/2\right) ^{2n}/\left( n!\right) ^2$
is the modified Bessel function, and $q$ is shown in Eq. (21). We define a
characteristic temperature $\widetilde{T}$ that is solution of equation $%
qe^{-\hbar \omega _1/2k_B\widetilde{T}}=1$. Then we obtain the specific heat
up to the order of $\left( \beta \hbar \Delta {\cal E}_0\right) ^2$ as 
\begin{eqnarray}
c &=&k_B\left( \beta \hbar \omega _1\right) ^2\frac{e^{\beta \hbar \omega _1}%
}{\left( e^{\beta \hbar \omega _1}-1\right) ^2}+\frac 12k_B\left( \beta
\hbar \Delta {\cal E}_0\right) ^2\left\{ \left[ 2\left( 1-e^{-\beta \hbar
\omega _1}\right) +4\left( \beta \hbar \omega _1\right) e^{-\beta \hbar
\omega _1}\right. \right.  \nonumber \\
&&\left. -\left( \beta \hbar \omega _1\right) ^2e^{-\beta \hbar \omega
_1}I_0\left( 2qe^{-\beta \hbar \omega _1/2}\right) \right] -q\left( \beta
\hbar \omega _1\right) \left[ \frac 12\left( 5e^{-3\beta \hbar \omega
_1/2}-e^{-\beta \hbar \omega _1/2}\right) \right.  \nonumber \\
&&\left. +4\left( e^{-\beta \hbar \omega _1/2}-e^{-3\beta \hbar \omega
_1/2}\right) \right] I_0^{\prime }\left( 2q_1e^{-\beta \hbar \omega
_1/2}\right) +q^2\left( \beta \hbar \omega _1\right) ^2\left( e^{-\beta
\hbar \omega _1/2}-e^{-3\beta \hbar \omega _1/2}\right)  \nonumber \\
&&\left. \times I_0^{\prime \prime }\left( 2qe^{-\beta \hbar \omega
_1/2}\right) \right\} ,  \eqnum{26}
\end{eqnarray}
for both integer and half-integer excess spins, where $I_0^{\prime }=-I_1$,
and $I_0^{\prime \prime }=I_2-I_1/x$. $I_\nu \left( x\right)
=\sum_{n=0}\left( -1\right) ^n\left( x/2\right) ^{2n+\nu }/n!\Gamma \left(
n+\nu +1\right) $, where $\Gamma $ is Gamma function. The results show that
the spin-parity effect will be lost at high temperatures. The specific heat
for integer excess spins is almost the same as that for half-integer excess
spins.

\section{MQC for trigonal, tetragonal and hexagonal symmetries}

In this section, we will apply the method in Sec. III to study resonant
quantum tunneling of the N\'{e}el vector in AFM particles with trigonal,
tetragonal and hexagonal crystal symmetry. For the trigonal symmetry, 
\begin{equation}
E\left( \theta ,\phi \right) =K_1\cos ^2\theta -K_2\sin ^3\theta \cos \left(
3\phi \right) +E_0,  \eqnum{27}
\end{equation}
where $K_1\gg K_2>0$. The energy minima of this system are at $\theta =\pi
/2 $, and $\phi =0$, $2\pi /3$, $4\pi /3$, and other energy minima repeat
the three states with period $2\pi $. The spin tunneling problem can be
mapped onto a problem of one-dimensional motion by integrating out the small
fluctuations of $\theta $ about $\pi /2$, and for this case $U\left( \phi
\right) =2\left( K_2V/\hbar \right) \sin ^2\left( 3\phi /2\right) $. Now $%
U\left( \phi \right) $ is periodic with period $2\pi /3$, and there are
three minima in the entire region $2\pi $. The periodic instanton
configuration with an energy $E>0$ is $\sin ^2\left( \frac 32\phi _p\right)
=1-k^2$sn$^2\left( \omega _2\tau ,k\right) $, where $k=\sqrt{\left(
n_1^2-1\right) /n_1^2}$, $\omega _2=3\frac V{\hbar S}\sqrt{\frac{K_2J}{1+%
\frac 12\left( \frac J{K_1}\right) \left( \frac sS\right) ^2}}$, and $n_1=%
\sqrt{2K_2V/\hbar E}>1$. The low energy limit of this periodic instanton
configuration agrees well with the vacuum instanton solution obtained in
Ref. 18. The associated classical action is ${\cal S}_p=W+2E\beta $, with 
\begin{equation}
W=\frac 83S\sqrt{\frac{K_2}J\left[ 1+\frac 12\left( \frac J{K_1}\right)
\left( \frac sS\right) ^2\right] }\left[ E\left( k\right) -\left(
1-k^2\right) K\left( k\right) \right] .  \eqnum{28}
\end{equation}
The general formula Eq. (4) gives the low-lying energy shift of $n$th
excited state as 
\begin{eqnarray}
\hbar \Delta {\cal E}_n^{tri} &=&\frac 3{\sqrt{2\pi }n!}\frac{\sqrt{K_2J}V}{S%
\sqrt{1+\frac 12\left( \frac J{K_1}\right) \left( \frac sS\right) ^2}}\left( 
\frac{32}3S\sqrt{\frac{K_2}J\left[ 1+\frac 12\left( \frac J{K_1}\right)
\left( \frac sS\right) ^2\right] }\right) ^{n+1/2}  \nonumber \\
&&\times \exp \left( -\frac 83S\sqrt{\frac{K_2}J\left[ 1+\frac 12\left( 
\frac J{K_1}\right) \left( \frac sS\right) ^2\right] }\right) .  \eqnum{29}
\end{eqnarray}
The periodic potential $U\left( \phi \right) $ can be viewed as a
superlattice with lattice constant $2\pi /3$ and total length $2\pi $, and
the Bloch theorem then gives the energy level spectrum of $n$th excited
state ${\cal E}_n^{tri}=\left( n+1/2\right) \omega _2$ as $E_n^{tri}={\cal E}%
_n^{tri}-2\Delta {\cal E}_n^{tri}\cos \left[ \left( S_{tot}+\xi \right) 2\pi
/3\right] $, where $\xi =-1,0,1$ in the first Brillouin zone. It is easy to
show that the low-lying energy level spectrum is $\hbar {\cal E}%
_n^{tri}-2\hbar \Delta {\cal E}_n^{tri}$, and $\hbar {\cal E}_n^{tri}+\hbar
\Delta {\cal E}_n^{tri}$ for integer excess spins, the latter being doubly
degenerate. While the level spectrum is $\hbar {\cal E}_n^{tri}-\hbar \Delta 
{\cal E}_n^{tri}$, and $\hbar {\cal E}_n^{tri}+2\hbar \Delta {\cal E}%
_n^{tri} $ for half-integer excess spins, the former being doubly degenerate.

For the tetragonal symmetry, the magnetocrystalline anisotropy energy is 
\begin{equation}
E\left( \theta ,\phi \right) =K_1\cos ^2\theta +K_2\sin ^4\theta
-K_2^{\prime }\sin ^4\theta \cos \left( 4\phi \right) +E_0,  \eqnum{30}
\end{equation}
where $K_1\gg K_2$, $K_2^{\prime }>0$. The energy minima are at $\theta =\pi
/2$, and $\phi =0$, $\pi /2$, $\pi $, $3\pi /2$, and other energy minima
repeat the four states with period $2\pi $. The problem can be mapped onto a
problem of particle moving in one-dimensional potential $U\left( \phi
\right) =2\left( K_2^{\prime }V/\hbar \right) \sin ^2\left( 2\phi \right) $
by integrating out the small fluctuations of $\theta $ about $\pi /2$. Now $%
U\left( \phi \right) $ is periodic with period $\pi /2$, and there are four
minima in the entire region $2\pi $. The periodic instanton configuration
with an energy $E>0$ is $\sin ^2\left( 2\phi _p\right) =1-k^2$sn$^2\left(
\omega _3\tau ,k\right) $, where $k=\sqrt{\left( n_1^2-1\right) /n_1^2}$, $%
\omega _3=4\frac V{\hbar S}\sqrt{\frac{K_2^{\prime }J}{1+\frac 12\left( 
\frac J{K_1}\right) \left( \frac sS\right) ^2}}$, and $n_1=\sqrt{%
2K_2^{\prime }V/\hbar E}>1$. The associated classical action is ${\cal S}%
_p=W+2E\beta $, with 
\begin{equation}
W=2S\sqrt{\frac{K_2^{\prime }}J\left[ 1+\frac 12\left( \frac J{K_1}\right)
\left( \frac sS\right) ^2\right] }\left[ E\left( k\right) -\left(
1-k^2\right) K\left( k\right) \right] .  \eqnum{31}
\end{equation}
The low-lying energy shift of $n$th excited state is 
\begin{eqnarray}
\hbar \Delta {\cal E}_n^{te} &=&\frac{2^{3/2}}{\sqrt{\pi }n!}\frac{\sqrt{%
K_2^{\prime }J}V}{S\sqrt{1+\frac 12\left( \frac J{K_1}\right) \left( \frac sS%
\right) ^2}}\left( 8S\sqrt{\frac{K_2^{\prime }}J\left[ 1+\frac 12\left( 
\frac J{K_1}\right) \left( \frac sS\right) ^2\right] }\right) ^{n+1/2} 
\nonumber \\
&&\times \exp \left( -2S\sqrt{\frac{K_2^{\prime }}J\left[ 1+\frac 12\left( 
\frac J{K_1}\right) \left( \frac sS\right) ^2\right] }\right) .  \eqnum{32}
\end{eqnarray}
Now $U\left( \phi \right) $ can be viewed as a superlattice with lattice
constant $\pi /2$ and total length $2\pi $, and the Bloch theorem gives the
energy level spectrum of $n$th excited state ${\cal E}_n^{te}=\left(
n+1/2\right) \omega _3$ as $E_n^{te}={\cal E}_n^{te}-2\Delta {\cal E}%
_n^{te}\cos \left[ \left( S_{tot}+\xi \right) \pi /2\right] $, where $\xi
=-1,0,1,2$ in the first Brillouin zone. Then the low-lying energy level
spectrum is $\hbar {\cal E}_n^{te}\pm 2\hbar \Delta {\cal E}_n^{te}$, and $%
\hbar {\cal E}_n^{te}$ for integer excess spins, the latter being doubly
degenerate. While the level spectrum is $\hbar {\cal E}_n^{te}\pm \sqrt{2}%
\hbar \Delta {\cal E}_n^{te}$ with doubly degenerate for half-integer excess
spins.

For the case of hexagonal symmetry, 
\begin{equation}
E\left( \theta ,\phi \right) =K_1\cos ^2\theta +K_2\sin ^4\theta +K_3\sin
^6\theta -K_3^{\prime }\sin ^6\theta \cos \left( 6\phi \right) +E_0, 
\eqnum{33}
\end{equation}
where $K_1\gg K_2,K_3,K_3^{\prime }>0$. The easy directions are at $\theta
=\pi /2$, and $\phi =0$, $\pi /3$, $2\pi /3$, $\pi $, $4\pi /3$, $5\pi /3$,
and other energy minima repeat the six states with period $2\pi $. For the
present case, $U\left( \phi \right) =2\left( K_3^{\prime }V/\hbar \right)
\sin ^2\left( 3\phi \right) $ is periodic with period $\pi /3$, and there
are six minima in the entire region $2\pi $. The periodic instanton
configuration at a given energy $E>0$ is $\sin ^2\left( 3\phi _p\right)
=1-k^2$sn$^2\left( \omega _4\tau ,k\right) $, where $k=\sqrt{\left(
n_1^2-1\right) /n_1^2}$, $\omega _4=6\frac V{\hbar S}\sqrt{\frac{K_3^{\prime
}J}{1+\frac 12\left( \frac J{K_1}\right) \left( \frac sS\right) ^2}}$, and $%
n_1=\sqrt{2K_3^{\prime }V/\hbar E}>1$. Correspondingly, the classical action
is ${\cal S}_p=W+2E\beta $, with 
\begin{equation}
W=\frac 43S\sqrt{\frac{K_2^{\prime }}J\left[ 1+\frac 12\left( \frac J{K_1}%
\right) \left( \frac sS\right) ^2\right] }\left[ E\left( k\right) -\left(
1-k^2\right) K\left( k\right) \right] ,  \eqnum{34}
\end{equation}
and the low-lying energy shift of $n$th excited state is 
\begin{eqnarray}
\hbar \Delta {\cal E}_n^{he} &=&\frac{3\times 2^{1/2}}{\sqrt{\pi }n!}\frac{%
\sqrt{K_3^{\prime }J}V}{S\sqrt{1+\frac 12\left( \frac J{K_1}\right) \left( 
\frac sS\right) ^2}}\left( \frac{16}3S\sqrt{\frac{K_3^{\prime }}J\left[ 1+%
\frac 12\left( \frac J{K_1}\right) \left( \frac sS\right) ^2\right] }\right)
^{n+1/2}  \nonumber \\
&&\times \exp \left( -\frac 43S\sqrt{\frac{K_3^{\prime }}J\left[ 1+\frac 12%
\left( \frac J{K_1}\right) \left( \frac sS\right) ^2\right] }\right) . 
\eqnum{35}
\end{eqnarray}
Now $U\left( \phi \right) $ can be regarded as a one-dimensional
superlattice with lattice constant $\pi /3$. By applying the Bloch theorem
and the tight-binding approximation, we obtain the energy level spectrum of $%
n$th excited state ${\cal E}_n^{he}=\left( n+1/2\right) \omega _4$ as $%
E_n^{he}={\cal E}_n^{he}-2\Delta {\cal E}_n^{he}\cos \left[ \left(
S_{tot}+\xi \right) \pi /3\right] $, where $\xi =-2,-1,0,1,2,3$. If the
excess spin $s$ is an integer, the low-lying energy level spectrum is $\hbar 
{\cal E}_n^{he}\pm 2\hbar \Delta {\cal E}_n^{he}$, and $\hbar {\cal E}%
_n^{he}\pm \hbar \Delta {\cal E}_n^{he}$, the latter two levels being doubly
degenerate. If $s$ is a half-integer, the level spectrum is $\hbar {\cal E}%
_n^{he}\pm \sqrt{3}\hbar \Delta {\cal E}_n^{he}$, and $\hbar {\cal E}_n^{he}$%
, all three levels being doubly degenerate.

In brief, the low-lying energy level spectrum for trigonal, tetragonal and
hexagonal symmetry are found to depend on the parity of the excess spins of
AFM particles distinctly, resulting from the phase interference between
topologically different tunneling paths. The structure of low-lying
tunneling level spectrum for the trigonal, tetragonal, or hexagonal symmetry
is found to be much more complex than that for the biaxial symmetry. The
low-lying energy level spectrum can be nonzero even if the excess spin is a
half-integer for the trigonal, tetragonal, or hexagonal symmetry. The
results of AFM nanoparticles with general structure of magnetocrystalline
anisotropy will be helpful for experimental test.

\section{Conclusions}

In summary, we have studied the topological phase interference effects in
the model for mesoscopic AFM particles with uncompensated excess spin for
the more general structure of magnetic anisotropy, such as biaxial,
trigonal, tetragonal, and hexagonal crystal symmetries. The low-lying tunnel
splittings between $n$th degenerate excited states of neighboring wells are
evaluated with the help of the periodic instanton method, and the energy
level spectrum is obtained by applying the Bloch theorem and the
tight-binding approximation in one-dimensional periodic potential. This is
the first complete study, to our knowledge, of spin-phase interference
between excited-level tunneling paths in AFM particles with general
structure of magnetocrystalline anisotropy, which will be useful for
experimental check.

One important conclusion is that for all the four kinds of crystal
symmetries, the low-lying energy level spectrum for integer excess spins is
significantly different from that for half-integer excess spins, resulting
from the phase interference between topologically distinct tunneling paths.
For AFM particles with simple biaxial symmetry, which has two degenerate
easy directions in the basal plane (i.e., the double-well system), the
tunnel splitting is suppressed to zero for half-integer excess spins due to
the destructive phase interference between topologically different tunneling
paths connecting the same initial and final states. However, the structure
of low-lying tunneling level spectrum for the trigonal, tetragonal, or
hexagonal symmetry is found to be much more complex than that for the
biaxial symmetry. The low-lying energy level spectrum can be nonzero even if
the excess spin is a half-integer for the trigonal, tetragonal, or hexagonal
symmetry. Our analytical study provides a nontrivial generalization of
Kramers degeneracy for double-well system to coherently spin tunneling at
ground states as well as low-lying excited states for AFM systems with $m$%
-fold rotational symmetry around the $\widehat{z}$ axis. Note that these
spin-parity effects are of topological origin, and therefore are independent
of the magnitude of excess spins of AFM particles, the shape of the soliton
and the tunneling potential. One can easily show that the heat capacity of
low-lying magnetic tunneling states depends significantly on the parity of
excess spins for AFM particles with different symmetries at sufficiently low
temperatures, providing a possible experimental method to examine the
theoretical results on topological phase interference effects. Our results
presented here should be useful for a quantitative understanding on the
topological phase interference or spin-parity effects in resonant quantum
tunneling in single-domain AFM particles with different symmetries.

Over the past years a lot of experimental and theoretical works were
performed on the spin tunneling in molecular Mn$_{12}$-Ac\cite{Mn12} and Fe$%
_8$\cite{Fe8} clusters having a collective spin state $S=10$ (in this paper $%
S=10^3$-$10^4$). More recently, Wernsdorfer and Sessoli\cite{Wernsdorfer}
have measured the tunnel splittings in the molecular Fe$_8$ clusters, and
have found a clear oscillation of the tunnel splitting with the field along
hard axis, which is a direct evidence of the role of the Berry phase in the
spin dynamics of these molecules. Further experiments should focus on the
level quantization of collective spin states of $S=10^2$-$10^4$ and their
quantum spin phases. The theoretical calculations performed in this paper
can be extended to the FM and AFM particles in a magnetic field. Similar
spin-phase interference effects observed in Fe$_8$ cluster should be found
in single-domain AFM nanoparticles in a magnetic field. Work along this line
is still in progress. With current technology and fast progress on this
field, our study on spin-phase interference and resonant quantum coherence
effects in AFM nanoparticles should be experimentally testable in the near
future.

\section*{Acknowledgments}

R. L. are grateful to Dr. S. -P. Kou, Prof. Z. Xu, Prof. J. -Q. Liang and
Prof. F. -C. Pu for stimulating discussions. R. L. and J. -L. Z. would like
to thank Prof. W. Wernsdorfer and Prof. R. Sessoli for providing their paper
(Ref. 19). The financial support from NSF-China (Grant No. 19974019) and the
state key project of fundamental research is gratefully acknowledged.


\begin{references}
\bibitem{Leggett}  A. O. Caldeira and A. J. Leggett, Ann. Phys. {\bf 149},
374 (1983); A. J. Leggett {\it et al.}, Rev. Mod. Phys. {\bf 59}, 1 (1987).

\bibitem{review}  For a review, see {\it Quantum Tunneling of Magnetization}%
, edited by L. Gunther and B. Barbara (Kluwer, Dordrecht, 1995); and E. M.
Chudnovsky and J. Tejada, {\it Macroscopic Quantum Tunneling of the Magnetic
Moment} (Cambridge University Press, 1997).

\bibitem{Awschalom}  D. D. Awschalom, J. F. Smyth, G. Grinstein, D. P.
DiVincenzo, and D. Loss, Phys. Rev. Lett. {\bf 68}, 3092 (1992), S. Gider,
D. D. Awschalom, T. Douglas, and M. Chaparala, Science {\bf 268}, 77 (1995);
J. Tejada, X. X. Zhang, E. del Barco, J. M. Hernandez, and E. M. Chudnovsky,
Phys. Rev. Lett. {\bf 79}, 1754 (1997).

\bibitem{Tejade}  J. Tejade, E. M. Chudnovsky, E. del Barco, J. M.
Hernandez, and T. P. Spiller, cond-mat/0009432 and references therein.

\bibitem{Fradkin}  E. Fradkin, {\it Field Theories of Condensed Matter
Systems} (Addison-Wesley, Reading, MA, 1992), Chap. 5.

\bibitem{Loss(92) and Delft (92)}  D. Loss, D. P. DiVincenzo, and G.
Grinstein, Phys. Rev. Lett. {\bf 69}, 3232 (1992); J. V. Delft and G. L.
Henley, Phys. Rev. Lett. {\bf 69}, 3236 (1992); Phys. Rev. B {\bf 48}, 965
(1993).

\bibitem{Garg (93) and Loss (95)(96)}  A. Garg, Europhys. Lett. {\bf 22},
205 (1993); H. B. Braun and D. Loss, Europhys. Lett. {\bf 31}, 555 (1995);
Phys. Rev. B {\bf 53}, 3237 (1996).

\bibitem{Barnes}  S. E. Barnes, Advances in Physics {\bf 30}, 801 (1981); J.
Phys.: Condens. Matter {\bf 6}, 719 (1994); {\bf 10}, L665 (1998);
cond-mat/9710302; cond-mat/9901219; cond-mat/9907257; S. E. Barnes, R.
Ballou, and J. Strelen, Phys. Rev. Lett. {\bf 79}, 289 (1997).

\bibitem{Bogachek and Krive}  E. N. Bogachek and I. V. Krive, Phys. Rev. B 
{\bf 46}, 14559 (1992).

\bibitem{AFM}  E. M. Chudnovsky, J. Magn. Magn. Mater. {\bf 140-144}, 1821
(1995); I. V. Krive and O. B. Zaslawski, J. Phys.: Condens. Matter {\bf 2},
9457 (1990); J. M. Duan and A. Garg, J. Phys.: Condens. Matter {\bf 7}, 2171
(1995).

\bibitem{Levine}  G. Levine and J. Howard, Phys. Rev. Lett. {\bf 75}, 4142
(1995).

\bibitem{Loss (97)(98)}  A. Chiolero and D. Loss, Phys. Rev. B {\bf 56}, 738
(1997); Phys. Rev. Lett. {\bf 80}, 169 (1998).

\bibitem{Chudnovsky93}  E. M. Chudnovsky and D. P. DiVincenzo, Phys. Rev. B 
{\bf 48}, 10548 (1993).

\bibitem{Garg95}  A. Garg, Phys. Rev. B {\bf 51}, 15161 (1995).

\bibitem{Wang}  Xiao-Bing Wang and Fu-Cho Pu, J. Phys.: Condens. Matter {\bf %
8}, L541 (1996).

\bibitem{Lu1}  Rong L\"{u}, Jia-Lin Zhu, Xi Chen, and Lee Chang, Euro. Phys.
J B{\bf 3},35 (1998); Rong L\"{u}, Jia-Lin Zhu, Xiao-Bing Wang, and Lee
Chang, Phys. Rev. B {\bf 58}, 8542 (1998).

\bibitem{Liang1}  J. -Q. Liang, Y.-B. Zhang, H. J. W. M\"{u}ller-Kirsten,
Jian-Ge Zhou, F. Zimmerschied, and F.-C. Pu, Phys. Rev. B {\bf 57}, 529
(1998); S. P. Kou, J. Q. Liang, Y. B. Zhang, and F. C. Pu, Phys. Rev. B {\bf %
59}, 11792 (1999).

\bibitem{Wang2}  Xiao-Bing Wang and Fu-Cho Pu, J. Phys.: Condens. Matter 
{\bf 9}, 693 (1997).

\bibitem{Lu2}  Jia-Lin Zhu, Rong L\"{u}, Xiao-Bing Wang, Xi Chen, Lee Chang,
and Fu-Cho Pu, Eur. Phys. J. {\bf B4}, 223 (1998); Rong L\"{u}, Jia-Lin Zhu,
Jian Wu, Xiao-Bing Wang, and Lee Chang, Phys. Rev. B {\bf 60}, 3534 (1999).

\bibitem{Wernsdorfer}  W. Wernsdorfer and R. Sessoli, Science {\bf 284}\-,
133 (1999).

\bibitem{Garanin}  M. N. Leuenberger and D. A. Garanin, cond-mat/9810156.

\bibitem{Garg}  A. Garg, Phys. Rev. Lett. {\bf 83}, 4385 (1999);
cond-mat/9907197; cond-mat/9907198; Phys. Rev. Lett. {\bf 81}, 1513 (1998).

\bibitem{Chiorescu}  I. Chiorescu, W. Wernsdorfer, A. M\"{u}ller, H. B\"{o}%
gge, and B. Barbara, Phys. Rev. Lett. {\bf 84}, 3454 (2000); W. Wernsdorfer,
R. Sessoli, A. Caneschi, D. Gatteschi, and A. Cornia, Europhys.Lett. {\bf 50}%
, 552 (2000).

\bibitem{Liang2}  J.-Q. Liang, H. J. W. M\"{u}ller-Kirsten, D. K. Park, and
F.-C. Pu, cond-mat/0001142.

\bibitem{Kittle}  C. Kittle, Rev. Mod. Phys. {\bf 21}, 552 (1949); E. P.
Wohlfarth, Proc. Phys. Soc. (London) A {\bf 65}, 1053 (1952); C. Herring,
Phys. Rev. {\bf 85}, 1003 (1952).

\bibitem{Coleman}  S. Coleman, {\it Aspects of Symmetry} (Cambridge
University Press, Cambridge, England, 1985), Chap.7.

\bibitem{Liang3}  J. -Q. Liang and H. J. W. M\"{u}ller-Kirstein, Phys. Rev.
D {\bf 46}, 4685 (1992); {\bf 50}, 6519 (1994); {\bf 51}, 718 (1995).

\bibitem{Landau}  L. D. Landau and E. M. Lifshita, {\it Quantum Mechanics}
3rd edition, (Pergamon, New York, 1977), Chap. VII, Sec. 50, Problem 3.

\bibitem{Garanin2}  D. A. Garanin, X. M. Hidalgo, and E. M. Chudnovsky,
Phys. Rev. B {\bf 57}, 13639 (1998), Sec. II. B.

\bibitem{Weiss}  U. Weiss and W. Haeffner, Phys. Rev. D {\bf 27}, 2916
(1983); H. K. Shepard, Phys. Rev. D {\bf 27}, 1288 (1983); A. Garg,
cond-mat/0003115.

\bibitem{Chudnovsky(PRL97)}  E. M. Chudnovsky and D. A. Garanin, Phys. Rev.
Lett. {\bf 79}, 4469 (1997).

\bibitem{Liang4}  J.-Q. Liang, H. J. W. M\"{u}ller-Kirsten, D. K. Park, and
F. Zimmerschied, Phys. Rev. Lett. {\bf 81}, 216 (1998); S. Y. Lee, H. J. W. M%
\"{u}ller-Kirsten, D. K. Park, and F. Zimmerschied, Phys. Rev. B {\bf 58},
5554 (1998).

\bibitem{Gorokhov}  D. A. Gorokhov and G. Blatter, Phys. Rev. B {\bf 57},
3586 (1998); {\bf 58}, 5486 (1998).

\bibitem{Park}  C. S. Park, S. -K. Yoo, D. K. Park, and D. -H. Yoon,
cond-mat/9807344; cond-matt/9902039; C. S. Park, S. -K. Yoo, and D. -H.
Yoon, cond-mat/9909217.

\bibitem{Weiss(book)}  U. Weiss, {\it Quantum Dissipative Systems,} (World
Scientific, 1999).

\bibitem{spin-dissipation}  A. Garg and G. -H. Kim, Phys. Rev. Lett. {\bf 63}%
, 2512 (1989); Phys. Rev. B {\bf 43}, 712 (1991); H. Simanjuntak, J. Low
Temp. Phys. {\bf 90}, 405 (1992); P. C. E. Stamp, Phys. Rev. Lett. {\bf 66},
2802 (1991); E. M. Chudnovsky, O. Iglesias, and P. C. E. Stamp, Phys. Rev. B 
{\bf 46}, 5392 (1992); A. Garg, Phys. Rev. Lett. {\bf 70}, 1541 (1993); {\bf %
74}, 1458 (1995); J. Appl. Phys. {\bf 76}, 6168 (1994); G. Tatara and H.
Fukuyama, Phys. Rev. Lett. {\bf 72}, 772 (1994).

\bibitem{Mn12}  R. Sessoli, D. Gatteschi, A. Caneschi, and M. A. Novak,
Nature {\bf 365}, 141 (1993); C. Paulsen and J. -G. Park, in {\it Quantum
Tunneling of Magnetization}, edited by L. Gunther and B. Barbara (Kluwer,
Dordrecht, 1995); M. A. Novak and R. Sessoli, in {\it Quantum Tunneling of
Magnetization}, edited by L. Gunther and B. Barbara (Kluwer, Dordrecht,
1995); J. M. Hernandez, X. X. Zhang, F. Luis, J. Bartolome, J. Tejada, and
R. Ziolo, Europhys. Lett. {\bf 35}, 301 (1996); L. Thomas, F. Lionti, R.
Ballou, D. Gatteschi, R. Sessoli, and B. Barbara, Nature (London) {\bf 383},
145 (1996); J. R. Friedman, M. P. Sarachik, J. Tejada, and R. Ziolo, Phys.
Rev. Lett. {\bf 76}, 3830 (1996); J. M. Hernandez, X. X. Zhang, F. Luis, J.
Tejada, J. R. Friedman, M. P. Sarachik, and R. Ziolo, Phys. Rev. B {\bf 55},
5858 (1997); F. Lionti, L. Thomas, R. Ballou, B. Barbara, A. Sulpice, R.
Sessoli, and D. Gatteschi, J. Appl. Phys. {\bf 81}, 4608 (1997); D. A.
Garanin and E. M. Chudnovsky, Phys. Rev. B {\bf 56}, 11102 (1997).

\bibitem{Fe8}  A.-L. Barra, P. Debrunner, D. Gatteschi, C. E. Schulz, and R.
Sessoli, Europhys. Lett. {\bf 35}, 133 (1996); C. Sangregorio, T. Ohm, C.
Paulsen, R. Sessoli, and D. Gatteschi, Phys. Rev. Lett. {\bf 78}, 4645
(1997).
\end{references}
\end{document}